\documentstyle[aps,prl,twocolumn,epsf,floats]{revtex}

\newcommand{\beq}{\begin{equation}}
\newcommand{\eeq}{\end{equation}}
\newcommand{\bea}{\begin{eqnarray}}
\newcommand{\eea}{\end{eqnarray}}

\newcommand{\veps}{\varepsilon}

\newcommand{\benn}{\begin{displaymath}}
\newcommand{\eenn}{\end{displaymath}}

\begin{document}

\title{\bf \Large Atomic--Molecular Condensates with
 Large Positive Scattering Length }

\author{ Aurel Bulgac$^1$ and Paulo F. Bedaque$^2$}

\address{$^1$Department of Physics, University of
Washington, Seattle, WA 98195--1560, USA}

\address{$^2$Lawrence--Berkeley Laboratory,
1 Cyclotron Road,  MS 70R0319,  Berkeley, CA 94720--8169}

\maketitle


\begin{abstract}

We show that in the limit of large and positive atom--atom scattering
length the properties of an atomic--molecular Bose--Einstein
Condensate (amBEC) are determined by an universal energy density
functional (EDF). We find that the optimal conditions for the
formation of a stable amBEC are in the regime where there are no
shallow trimers and the atom--dimer scattering length is negative and
comparable in magnitude with the atom--atom scattering length.  At
temperatures lower than $T_c$ the chemical potentials for the atoms
and molecules can be specified independently. Besides three--body
recombinations processes into dimers of large size, inelastic
processes involving the formation of deeply bound small size molecular
states are possible. These inelastic processes do not lead to an
efficient heating of the amBEC and can be used for its mostly
non--destructive monitoring.

\end{abstract}

\draft



The theoretical prediction \cite{feshbach_t}, along with the
experimental confirmation \cite{feshbach_e} of the fact that one can
manipulate relatively easily the interaction of slow atoms by means of
external fields, opened the doors to a long series of possible
experiments. Perhaps among the most spectacular experimental results
obtained lately are the creation of the so--called Bosenova
\cite{bosenova} and the evidence that something related to a mixed
atomic--molecular Bose--Einstein Condensate (BEC) was created
\cite{amBEC_e}, an object envisioned by theory several years before
\cite{amBEC_t,eddy,heizen}, see also Refs. \cite{kh,comment} for a
couple of theoretical analyses of this experiment.  In the present
work we identify a new universal regime, which appears in the case
when the two--body scattering length $a$ is positive ($a>0$) and
large, $a \gg r_0$, where $r_0$ is the effective range, appearing the
the low energy parameterization of the $s$--wave phase shift $k\cot
\delta(k) =-1/a + r_0k^2/2+\ldots $ Typically $r_0$ is of the order of
the so called van der Waals length $(C_6m/\hbar^2)^{1/4}$
\cite{pethick}.  The universality of this new regime is encoded in
only two parameters, which define the effective Hamiltonian (except
for the mass), one of them the two--body scattering length $a$ and the
other a three--body characteristic, which can be chosen in various
ways. As we will show below, in this regime, when $na^3\ll 1$, the
properties of a cold mixture of atoms and shallow dimers are described
by a very simple theoretical model. In the other limit, when $a<0$ and
$|a|\gg r_0$, there is another universal regime and a new class of
dilute quantum liquids exists \cite{boselets_1}. The basic ingredients
of the model are derived from the well known Efimov effect
\cite{efimov_1,efimov_2}, which pertains to the universal properties
of the three--body systems with large two--body scattering lengths.

In order to derive the energy density of an atomic--molecular gas let
us first consider the dilute regime where $nr_0^3\ll 1$, with $n$ a
typical atomic or molecular density and $a={\cal{O}}(r_0)$.  The
details of the potential are not important and the atom--atom
interactions can be described by a contact term
\beq\label{eq:Ha}
H_a = - \psi^\dagger_a \frac{\hbar^2
\bbox{\nabla}^2}{2m} \psi_a
+ \frac{1}{2}\lambda_{2}\psi^{\dagger\;2}_a\psi_a^2
+ \frac{1}{3}\lambda_{3}\psi^{\dagger\;3}_a\psi_a^3,
\eeq
where $\psi^\dagger_a, \psi_a$ are creation and annihilation operators
for atoms. The value of $\lambda_2$ is determined (after fixing a
short distance regulator) by the scattering length $a$.  $\lambda_3$
depends also on a genuinely three--body length scale, denoted below by
$a_3^\prime$, that can be determined only through the value of a
three--body observable. Terms with more derivatives or higher body
forces can be included, but their effect on a dilute gas is small as
long as the relevant momenta satisfy the condition $kr_0\ll 1$
\cite{r0,ly_etal,furnstahl,molecules}. The Hamiltonian in
Eq.~(\ref{eq:Ha}) contains, in principle, all the information
necessary to describe the system in the dilute regime, including the
molecular states/dimers with binding energy $\hbar^2/ma^2$ and radius
$\approx a$ when $a>0$. In the case $a\gg r_0$ we are interested in,
the presence of bound states within the regime of validity of
Eq.~(\ref{eq:Ha}) evidences the fact that perturbation theory breaks
down. Since particle--particle diagrams should be summed up to all
orders, a simple mean--field approximation is not legitimate as well
and two--particle correlations have to accounted for explicitly.
Thus, even though $H_a$ contains all the information needed to
describe molecular states, it is inconvenient to use it directly
\cite{furnstahl}.

\begin{figure}[tbh]
\begin{center}
\epsfxsize=6.0cm
\centerline{\epsffile{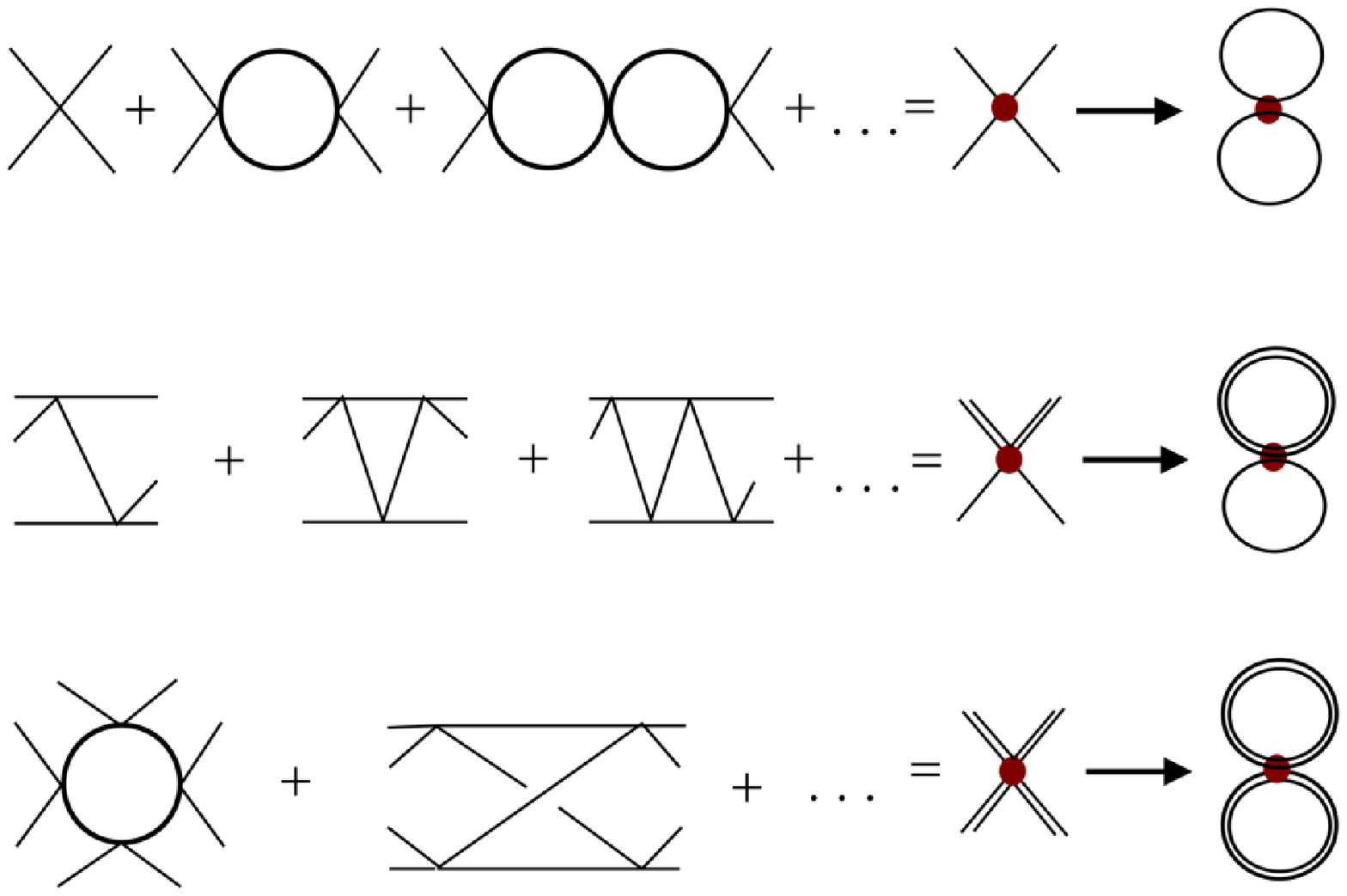}}
\end{center}

\caption{Matching between the 2--, 3-- and 4--particle amplitudes
computed with $H_a$ (left side of the equality) and $H_{am}$ (right
side). Only diagrams containing $\lambda_2$--vertices are shown. In a
description which includes explictly Feshbach molecules, each of these
vertices stand for a sum of two diagrams, a
two--atoms$\rightarrow$two--atoms diagram and a
two--atoms$\rightarrow$Feshbach molecule$\rightarrow$two--atoms
diagram. The effective vertices thus defined (right side) can then be
used to compute the ground state interaction energy in the leading
order terms in an $na^3$ expansion, which is given by the diagrams
after the arrows.}

\label{fig:diagrams}
\end{figure}

\begin{figure}[tbh]
\begin{center}
\epsfxsize=6.0cm
\centerline{\epsffile{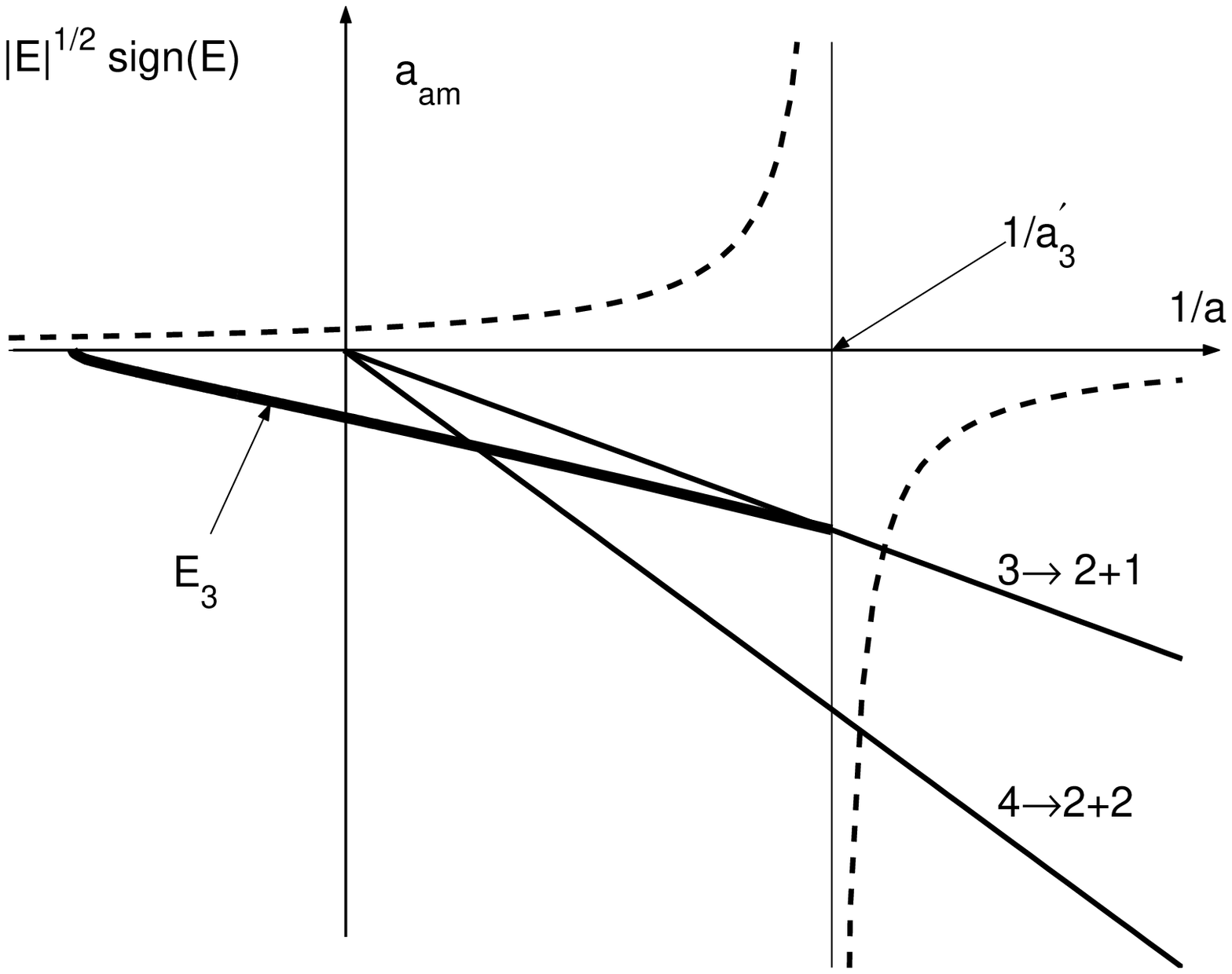}}
\end{center}

\caption{ The energy of an arbitrary Efimov state (thick solid line
labeled $E_3$) as a function of $1/a$. Following Efimov's argument we
use $|E|^{1/2}\mathrm{sign} (E)$ on the vertical axis instead of
energy.  The two straight lines labeled $3\rightarrow 2+1$ and
$4\rightarrow 2+2$ are the corresponding three-- and four--body
thresholds. The dashed line shows the qualitative behavior of the
scattering length $a_{am}$.}

\label{fig:levels3}
\end{figure}

Let us consider now a regime even more dilute, where $na^3\ll 1$. The
relevant typical momenta are of the order $p=\hbar k \approx
\sqrt{na^3}\hbar/a\ll \hbar/a \ll \hbar/r_0$ \cite{r0}.  We will argue
later that transitions altering the number of molecular states are
``very slow" in this regime and, consequently, for time scales shorter
than the transition rate, the number of atoms and molecules are
separately conserved. An appropriate effective Hamiltonian describing
the system in this regime is
\begin{eqnarray}\label{eq:Ham}
H_{am} &=&
- \psi^\dagger_a \frac{\hbar^2\bbox{\nabla}^2}{2m} \psi_a
- \psi^\dagger_m (\frac{\hbar^2\bbox{\nabla}_a^2}{4m}-\veps_2) \psi_m \\
  & + &
 \frac{1}{2}\lambda_{aa}\psi^{\dagger\;2}_a\psi_a^2+
  \lambda_{am}\psi^{\dagger}_a\psi^{\dagger}_m\psi_a\psi_m+
\frac{1}{2}\lambda_{mm}\psi^{\dagger\;2}_m\psi_m^2 .\nonumber
\end{eqnarray}
Both $H_a$ and $H_{am}$ are applicable in the regime $ka \ll kr_0 \ll
1$. $H_{am}$ has a great advantage over $H_a$ however. Perturbation or
mean--field theory is not valid for $H_a$ when $p\sim \hbar/a$, but it
is valid for $H_{am}$ since all non--perturbative physics occurring at
the scale $\sim a$ and leading to the formation of the bound state
(dimers of size $\sim a$) and to the Efimov effect (trimers of size
$\sim a$) is already neatly encapsulated in the constants
$\lambda_{aa}, \lambda_{am}$ and $\lambda_{mm}$. To make that more
explicit, let us determine $\lambda_{aa}, \lambda_{am}$ and
$\lambda_{mm}$ in terms of $\lambda_2$ and $\lambda_3$ by considering
zero momentum scattering of two atoms, atom--molecule and two
molecules.  Using $H_a$ these processes are described by an infinite
number of diagrams shown in Fig. \ref{fig:diagrams}, corresponding to
the solution of the Lippmann--Schwinger, Fadeev and Yakubovski
equations respectively
\cite{efimov_1,efimov_2,furnstahl,EFT_review}. The explicit solutions
of the Lippmann--Schwinger and Fadeev equations are known.  If we use
dimensional regularization \cite{hooft}, where divergences
proportional to powers of the momentum cutoff vanish, the same
processes are described using $H_{am}$ simply by the Born amplitude,
see the Fig. \ref{fig:diagrams}.  This leads to the identification
\bea
& & \lambda_{aa}= \frac{4\pi\hbar^2 a}{m},
\quad \varepsilon_2 = -\frac{\hbar^2}{ma^2}, \\
& & \lambda_{am}= \frac{3\pi\hbar^2a_{am}}{m}=\frac{3\pi\hbar^2 a}{m}
\left [ c_1 + c_2 \cot \left ( s_0 \ln \frac{a}{a_3^\prime }
  \right ) \right ],\label{lam} \\
& & \lambda_{mm}=\frac{2\pi \hbar^2a_{mm}}{m}=\frac{4\pi\hbar^2 a}{m} c_3,
\eea
where the values of the universal constants $s_0\approx 1.00624$,
$c_1\approx 1.46$ and $c_2\approx 2.15$ are known \cite{eric} and
$|\veps_2|$ is the true dimer binding energy.  $a_3^\prime>0$ is the
value of the two--body scattering length for which a trimer energy is
exactly at the atom--dimer threshold and this constant is system
dependent, see Fig. \ref{fig:levels3}. The specific value of
$a_3^\prime>0$ is determined by short--range three--particle physics,
when all three atoms are in a volume of order $r_0^3$. The functional
form of the atom--dimer scattering length was established by Efimov
\cite{efimov_2}. It is natural to expect also that the dimer--dimer
scattering length is of order $a$, as there are no known four--body
shallow states near the $4\rightarrow 2+2$ threshold and $a\approx
a_3^\prime$.  The value of $c_3$ has not yet been calculated for
bosons, but for fermions Randeria has shown at mean--field level that
$c_3=1$ \cite{mohit}. This result, in connection with Gribakin and
Flambaum's reasoning \cite{gribakin}, leads us to expect that in most
cases the dimer--dimer scattering length is positive. Notice that the
matching described above is performed in vacuum (length scales of
${\cal{O}}(a)$) and it is not modified by the many--body physics
(length scales of ${\cal{O}}(a/\sqrt{na^3})\gg {\cal{O}}(a)$).  The
fact that the dimensionally regulated $H_{am}$ is perturbative in the
$ka \ll 1$ regime, unlike $H_a$, is firmly established
\cite{furnstahl,EFT_review}.

The leading order contributions to the energy density functional are
given by the mean field approximation of $H_{am}$, see
Fig. \ref{fig:diagrams} ( NB, a mean--field approximation for $H_{a}$
would be highly inaccurate even at these low densities!),
\beq
{\mathcal{E}}= \frac{1}{2}\lambda_{aa} n_a^2 + \lambda_{am}n_an_m
+\frac{1}{2}\lambda_{mm}n_m^2 +\varepsilon_2 n_m ,\label{eq:edf}
\eeq
where $n_a, n_m$ are the atom and dimer densities (see below the
discussion on trimers contribution).  In terms of many--body diagrams
built using Eq.~(\ref{eq:Ha}) this expression corresponds to the exact
summation of the linear and quadratic terms in densities of {\bf all}
the two--particle, three--particle and four--particle diagrams,
computed with the initial Hamiltonian $H_a$.  Using well--known
arguments \cite{pethick,ly_etal}, it is straightforward to show that
higher order corrections, due to mean--field fluctuations, are
controlled by the diluteness parameter $\sqrt{na^3}\ll 1$.  This EDF
expression is remarkable, since, in the Efimov regime when $a \gg
r_0$, it is universal. Only the exact value of the parameter $c_3$ for
bosons is not explicitly known. Notice, however, that the condition
$na^3\ll 1$ precludes its use arbitrarily close to a Feshbach
resonance.  By changing the units of energy to $\hbar^2/ma^2$ and
length to $a$ this EDF acquires the following dimensionless and thus
universal form:
\beq
 {\mathcal{E}}= 2\pi  n_a^2 + \frac{3\pi\xi }{2} n_an_m
+2\pi c_3 n_m^2 - n_m,
\eeq
with $\xi=c_1+c_2\cot[s_0\ln (a/a_3^\prime)]$ tunable.

In the $a>0$ regime only the atom--dimer scattering length can change
dramatically from negative to positive values and even become
infinite. If $\lambda_{aa}>0$, $\lambda_{mm}>0$ and
$\lambda_{aa}\lambda_{mm}-\lambda_{am}^2>0$, then a mixed dilute
atom--molecule system is a metastable gas.  This can be seen as well
by computing the pressure
\beq
P=\frac{1}{2}\lambda_{aa} n_a^2 +\lambda_{am}n_an_m
+\frac{1}{2}\lambda_{mm}n_m^2.\label{eq:P}
\eeq
Phase separation occurs when $\lambda_{am} >
\sqrt{\lambda_{aa}\lambda_{mm}}$ and domain wall structures are
expected to appear \cite{dw}. A (meta)stable trimer phase might also
form if the (yet unknown) trimer--trimer interaction is repulsive.  
Another type of domain walls can appear in the opposite limit
\cite{son}. It is easy to see that there is no region where $P=0$ and
thus no stable liquid phase exists, unlike Refs. \cite{eddy}.  An
amBEC system would be in chemical equilibrium if
\beq
 (2\lambda_{aa}-\lambda_{am})n_a
+(2\lambda_{am}-\lambda_{mm})n_m
=\varepsilon_2.  \label{eq:mu}
\eeq
This type of chemical equilibrium can be established through the
reaction $A+A+A\rightarrow A+A_2$, which has a low rate for dilute
systems $ \propto \hbar a^4n_a^2/m$, see Refs. \cite{recomb} and
below. At temperatures $T< T_c\propto (na^3)^{2/3} \hbar^2/ma^2 \ll
|\veps_2|=\hbar^2/ma^2$, the rhs of the above chemical equilibrium
condition is parametrically much larger than the lhs when $na^3\ll 1$.
The reverse reaction $A_2+A\rightarrow A+A+A$ is therefore
exponentially suppressed, as it requires an activation energy larger
then the thermal energy.  Consequently, these type of collisions thus
lead to a very small heating (and very slow chemical equilibration of
the system) and its ultimate depletion \cite{recomb}. Since these two
reactions (and $A_2+A_2\leftrightarrow A_2 + A + A$, etc. as well) are
very slow in an amBEC, the chemical potentials for atoms and dimers
could be specified independently
$$
\mu_a= 
\lambda_{aa} n_a+\lambda_{am}n_m,
    \quad
\mu_m= 
\lambda_{am} n_a+\lambda_{mm}n_m
 +\varepsilon_2.
$$
In a trap one has to add the trap potentials for atoms and dimers in
the rhs of these two relations \cite{pethick}.

A number of other processes can occur in such a system.  It is
desirable to consider three--body inelastic collisions involving one
or more dimers in the initial state. Particularly interesting are
however the inelastic atom--dimer $A+A_2\rightarrow A^*+A^*_2$ (see
Ref. \cite{bh} for a rate calculation) and dimer--dimer
$A_2+A_2\rightarrow A^*+A^*+ A^*_2$ collisions, where $A^*$ and
$A_2^*$ stand for a fast atom (with its kinetic energy significantly
exceeding the thermal energy $3T/2$) and $A_2^*$ for a fast dimer
respectively (with a binding energy of the order $\hbar^2/mr_0^2\gg
\hbar^2/ma^2$). If the temperature of the system is significantly
lower then the dimer binding energy, i.e. $T<T_c \ll \hbar^2/ma^2$,
these inelastic collisions are not leading to a significant heating of
the system, but mostly to its depletion, in a manner similar to the
depletion in the BEC of metastable helium \cite{helium}. If an atom
and a dimer collide and this leads to the formation of a dimer with a
binding energy of order $\hbar^2/mr_0^2\gg \hbar^2/ma^2$ a relative
large amount of energy is released (NB $T<T_c\ll \hbar^2/ma^2 \ll
\hbar^2/mr_0^2$). It is conceivable that transitions to various final
bound states could be resolved and thus the spectrum of fast ejectiles
would be discrete. The momenta of the outgoing atom and of the deeply
bound dimer are large as well, of the order $\hbar/r_0\gg \hbar/a$,
and such fast atoms and dimers will interact relatively weakly with
the rest of the atoms and dimers.  While the slow/thermal atoms and
shallow dimers interact with each other with cross sections of the
order of $a^2$, a fast atom and a deeply bound and fast dimer will
interact with the amBEC with a cross section of
order $r_0^2\ll a^2$.  One can expect a similar behavior in a
dimer--dimer inelastic reaction. The fact that the products of these
inelastic processes are fast and weakly interacting objects can be
used for a non--destructive monitoring of the amBEC. The rates of
these inelastic processes are obviously controlled by the parameter
$r_0^3/a^3 \ll 1$, since in order to form a deeply bound dimer two
atoms in the shallow dimer have to change their separation from
distances of order $a$ to distances of order $r_0$.  One cannot help
here but to make a parallel with the similar role played by neutrinos
(which are also fast and weakly interacting products of inelastic
processes) in studying the interior of the Sun.

A possible final channel of a dimer--dimer collision could be a
trimer--atom final state if $a_{am}>0$.  If the trimer energy is close
to the atom--dimer threshold, then the atom--dimer cross section could
be significantly larger than the atom--atom and dimer--dimer cross
sections ($|a_{am}|\gg a,a_{mm}$).  If this is the case, one can expect
that the atom--dimer elastic collisions would be chiefly responsible
for the thermalization of the system. However, besides the elastic
channel there is also an inelastic channel present if $a_{am}>0$: a
dimer--dimer collision leading to a shallow trimer and an atom in the
final state. This process will lead to a relatively small release of
energy since the trimer energy is very close to the threshold, see
Fig. \ref{fig:levels3}. With good accuracy the ground state energy of
a trimer is given in this limit by
$\varepsilon_3\approx  \varepsilon_2 -3\hbar^2/4ma_{am}^2.$ \cite{efimov_1}
Moreover, the trimer under these conditions is basically a very large
two--body object of size $a_{am}\gg a$, made of a dimer and a
relatively loosely bound atom to it. In this limit the atoms and
dimers will phase separate, since $\lambda_{am} \gg
\sqrt{\lambda_{aa}\lambda_{mm}}$.  A pure trimer phase could appear
instead, iff the trimer--trimer interaction is repulsive.  When
$a_{am}>0$ and $\lambda _{am}\le \sqrt{\lambda_{aa}\lambda_{mm}}$
dimer--dimer collisions with a trimer in the final state will lead to
an energy release of the order of $\hbar^2/ma^2$, and thus to the
heating of the system, since the momenta of the emerging products are
of the order $\hbar/a$ in this case.  Consequently, the best
conditions to create a stable amBEC are expected when $a_{am}<0$ (no
shallow bound trimers and thus no $A_2+A_2\rightarrow A_3+A$
reactions) and $|a_{am}|< a$.  As a last note, there is essentially no
overlap between present results and those of Refs. \cite{others}.

Discussions with David B. Kaplan and Lan Yin are greatly appreciated.




\begin{thebibliography}{99}
\bibitem{r0} In medium the relevant momenta are determined by
estimating the quantum fluctuations of the mean--field and one thus
obtains that $p=\hbar k \approx \hbar \sqrt{na}$. Only as long as
$kr_0\approx ka \propto \sqrt{na^3} \ll 1$ one can replace the
strength of the contact interaction by the scattering length $a$.  The
accuracy of the mean--field approximation can be ascertained by
estimating the magnitude of quantum fluctuations to EDF, namely
$\propto n(na^3)^{4/3} \hbar^2/ma^2$ \cite{ly_etal}.
\bibitem{molecules} The typical Hamiltonian used to describe a system
of atoms near a Feshbach resonance \cite{amBEC_t,eddy,kh} includes
explicitly bound Feshbach molecules of size $\approx r_0$ in the
closed channel. The relative energy separation between this Feshbach
molecule and the two--atom threshold (open channel) is manipulated by
means of a magnetic field. The full two--atom wave function has at
small separations ${\cal{O}}(r_0)$ a component corresponding to two
atoms in a Feshbach molecule, which vanishes at large separations.
The existence of this contribution can be described accurately for
example by a boundary condition on the atom--atom wave function if
$kr_0 \ll 1$.  The explicit inclusion of a Feshbach molecule and its
coupling to the open two--atom channel (if treated exactly in the
regime $kr_0 \ll 1$) is no more accurate than using Eq. (\ref{eq:Ha}),
see also Refs. \cite{comment}, only more complicated, requiring
non--mean--field contributions and relying on a larger number of (not
independent) parameters.
\end{thebibliography}
\end{document}